\documentclass[reprint,aps,prl,amsmath,amssymb]{revtex4-1}
\usepackage{color}
\usepackage{amsmath}
\usepackage{bm}
\usepackage{graphicx}
\usepackage{multirow}
\usepackage{float}

\begin{document}

\newcommand{\bn}{\bm n}
\newcommand{\bp}{\bm{p}}   
\newcommand{\br}{\bm{r}}
\newcommand{\bk}{\bm{k}}
\newcommand{\bv}{\bm{v}}
\newcommand{\brho}{{\bm{\rho}}}
\newcommand{\bj}{{\bm j}}
\newcommand{\wk}{\omega_{\bm{k}}}
\newcommand{\nk}{n_{\bm{k}}}
\newcommand{\eps}{\varepsilon}
\newcommand{\la}{\langle}
\newcommand{\ra}{\rangle}
\newcommand{\be}{\begin{eqnarray}}
\newcommand{\ee}{\end{eqnarray}}
\newcommand{\intl}{\int\limits_{-\infty}^{\infty}}
\newcommand{\dE}{\delta{\cal E}^{ext}}
\newcommand{\SE}{S_{\cal E}^{ext}}
\newcommand{\dsp}{\displaystyle}
\newcommand{\phit}{\varphi_{\tau}}
\newcommand{\p}{\varphi}
\newcommand{\cL}{{\cal L}}
\newcommand{\dphi}{\delta\varphi}
\newcommand{\dbj}{\delta\bm{j}}
\newcommand{\dI}{\delta I}
\newcommand{\dph}{\delta\varphi}
\newcommand{\ua}{\uparrow}
\newcommand{\da}{\downarrow}
\newcommand{\ip}{\{i_{+}\}}
\newcommand{\im}{\{i_{-}\}}
\newcommand{\tp}{\tilde p}
\newcommand{\change}[1]{{#1}}

\title{Josephson linewidth in a resistively-shunted model with non-sinusoidal current-phase relation}

\author{K. E. Nagaev} 

\affiliation{Kotelnikov Institute of Radioengineering and Electronics, Mokhovaya 11-7, Moscow 125009, Russia}

\begin{abstract}
This is a detailed derivation of the equation for the voltage noise of a resistively shunted Josephson 
junction with non-sinusoidal current-phase relation, which was presented in Sov. Phys. JETP {\bf 67}, 579 (1988).
\end{abstract}

\maketitle

The Resistively Shunted Junction (RSJ) model \cite{McCumber68,Stewart68} has been used for analyzing the dynamics 
of Josephson junctions for more than five decades. Though it does not capture some essential microscopic effects
like multiple Andreev reflections, it is simple and allows obtaining analytical results in many cases. Initially
it involved a sinusoidal superconducting current--phase relation (CPR) characteristic of tunnel junctions and short
metallic microbridges near the critical temperature, but then it became obsolete for these systems because a
microscopic description of their dynamics was developed. However more recently, new classes of superconducting weak 
links have emerged for which a microscopic description of non-stationary effects is still missing. A good example are superconductor--graphene--superconductor
junctions. Though they exhibit a strongly non-sinusoidal CPR \cite{Titov06}, the RSJ model is 
widely used for describing the dynamic effects in them \cite{Lee18}. 

The RSJ model not only gives the average current--voltage characteristic of a superconducting weak link, but also
may be used for calculating the spectrum of voltage noise when it is in the resistive state. First it was done in
\cite{Likharev72} for the sinusoidal CPR and then this result was extended to the case of its arbitrary shape 
\cite{Kogan88}. However the expression for the noise in the RSJ model with arbitrary CPR was given in this paper 
without derivation, and here  we present it.

We start with the equation of resistively shunted Josephson contact biased by a fixed current $I >I_c$ and neglect 
the effects of electrostatic capacitance. The equation of motion for the superconducting phase $\phi$ is of the
form
\be
 I = \omega_c^{-1} I_c\,\dot{\phi} + I_s(\phi) + \delta I(t),
 \label{RSJ}
\ee
where $\omega_c = (2e/\hbar)RI_c$ is the characteristic frequency, $R$ is the shunting resistance, and $\delta I(t)$ 
is a random thermal-noise current generated by the resistance $R$ at temperature $T$ with a correlation function
\be
 \la \delta I(t)\,\delta I(t')\ra = 2\,\frac{T}{R}\,\delta(t-t').
 \label{sources}
\ee
The phase dependence of supercurrent  $I_s(\phi)$ may be an arbitrary $2\pi$-periodic function. 

First of all, we set $\delta I=0$ and calculate the time-averaged voltage 
\be
 \overline{U} = \frac{\hbar}{2e}\,\overline{\dot\phi}
    \equiv \frac{\hbar}{2e}\,\omega_J.
 \label{U_1}
\ee
The period of Josephson oscillations may be calculated as
\be
 \frac{2\pi}{\omega_J} = \int_0^{2\pi} \frac{d\phi}{\dot\phi}
 = \frac{I_c}{\omega_c} \int_0^{2\pi} \frac{d\phi}{I - I_s(\phi)},
 \label{T_J}
\ee
hence
\be
 \overline{U} = 2\pi R \left[ \int_0^{2\pi} \frac{d\phi}{I - I_s(\phi)} \right]^{-1}.
 \label{U_2}
\ee
A direct differentiation of this equation with respect to $I$ immediately gives us
\be
 R_d = \frac{d\overline{U}}{dI} = \frac{\overline{U}^2}{2\pi R}\,
 \int_0^{2\pi} \frac{d\phi}{[I - I_s(\phi)]^2}
 \label{Rd}
\ee
and
\be
 \frac{dR_d}{dI} = 2\,\frac{R_d^2}{\overline{U}}
 - \frac{1}{\pi}\,\frac{\overline{U}^2}{R}\,\int_0^{2\pi} \frac{d\phi}{[I - I_s(\phi)]^3}.
 \label{dRdI}
\ee

The presence of random current $\delta I$ in Eq. \eqref{RSJ} may be taken into account by replacing
its argument by $t + \delta\tau$, where $\delta\tau$ is a random additive. In general, $\delta\tau$ is not small 
because it piles up during a long time, but its change during one period of Josephson oscillations is small. To obtain 
an equation for it, one may express the time derivative of $\phi$ from Eq. \eqref{RSJ} and invert it to obtain
\be
 \frac{d}{d\phi}\,(t+\delta\tau) = \frac{\omega_c^{-1} I_c}{I - I_s(\phi) - \delta I(t+\delta\tau)}.
 \label{dt/dphi}
\ee
Linearizing Eq. \eqref{dt/dphi} with respect to $\delta I$ results in the equation
\be
 \frac{d\delta\tau}{d\phi} = \frac{I_c}{\omega_c}\,\frac{\delta I(t+\delta\tau)}{[I - I_s(\phi)]^2}.
 \label{dtau}
\ee

With account taken of $\delta I$, the time dependence of voltage takes up the form
\begin{multline}
 U(t+\delta\tau) = (1 + \delta\dot{\tau})
 \\ \times
 \left\{
   \overline{U} + {\rm Im}\sum_{n>0} U_n \exp[i\,(\omega_nT + \delta\theta_n)]
 \right\},
 \label{U(t)}
\end{multline}
where $\omega_n = n\,\omega_J$ and $\delta\theta_n = n\,\omega_J \delta\tau$ are the phase shifts of the corresponding 
harmonics of voltage. It follows from Eq. \eqref{dtau} that the correlation time of $\delta\dot{\tau}$ is zero, so the
first factor in the right-hand side of Eq. \eqref{U(t)} affects only the lower part of oscillation line, while its upper 
part is determined by the diffusion coefficient of the phase shift $\delta\theta_n$. This diffusion coefficient may be 
calculated as
\be
 D_n = \frac{1}{2}\,\frac{\la\Delta\theta_n^2\ra}{2\pi/\omega_J},
 \label{Dn}
\ee
where $\Delta\theta_n$ is the change in $\delta\theta_n$ during the period of Josephson oscillations $2\pi/\omega_J$.
This change equals 
\begin{multline}
 \Delta\theta_n = n \omega_J \int_0^{2\pi} d\phi\,\frac{d\delta\tau}{d\phi}
 \\=
 n\,\frac{\omega_J}{\omega_c}\,I_c \int_0^{2\pi} d\phi\,\frac{\delta I(t)}{[I - I_s(\phi)]^2},
 \label{Dtheta}
\end{multline}
and $\la\Delta\theta_n^2\ra$ is obtained by averaging the product of two instances of Eq. \eqref{Dtheta}.
%
%
The next step is to rewrite the correlation function Eq. \eqref{sources} in terms of $\phi$ and $\phi'$ instead
of $t$ and $t'$ using the equality
\be
 \delta(t-t') = \dot{\phi}\,\delta(\phi-\phi'), 
\label{t-t'}
\ee
where $\dot{\phi}$ is obtained by setting $\delta I=0$ in Eq. \eqref{RSJ}. As a result, one obtains
\be
 \la\Delta\theta_n^2\ra = 2\,n^2\,\frac{T}{R}\,\frac{I_c \omega_J^2}{\omega_c} 
 \int_0^{2\pi} \frac{d\phi}{[I - I_s(\phi)]^3}.
 \label{D2}
\ee
The linewidth of harmonic oscillations subject to phase diffusion is $\Gamma_n = D_n/2$, and the spectral density of voltage is 
related with it 
by 
\be
 S_U(0) = \frac{\hbar^2}{e^2}\,\Gamma_1.
 \label{SU-1}
\ee
This relation is based only on the Wiener--Khinchin theorem and the Josephson relation $U=\hbar\dot{\phi}/2e$. Its 
derivation does not involve any assumptions on the current--phase relation and is independent of the microscopic model
\cite{Kogan-book}.
The integral over $\phi$ in the resulting expression may be eliminated by means of Eq. \eqref{dRdI}, and finally one arrives 
at the equation
\be
 S_U(0) = 4\,\frac{T}{R} \left( R_d^2 - \frac{1}{2}\,\overline{U}\,\frac{dR_d}{dI} \right),
 \label{SU-2}
\ee
which is precisely Eq. (51) from \cite{Kogan88}. As this equation contains only the parameters that can be extracted
directly from the current--voltage characteristic of the Josephson junction, it may serve as a convenient test for the
validity of the RSJ model for it.

I am grateful to Abigail Wessels for reminding me of this result.

\end{document}